# Focused Electron and X-ray Beam Crosslinking in Liquids for Nanoscale Hydrogels 3D Printing and Encapsulation

Tanya Gupta[1,2], Evgheni Strelcov[1,2], Glenn Holland[1], Joshua Schumacher[1], Yang Yang[1], Mandy Esch[1], Vladimir Aksyuk[1], Patrick Zeller[3], Matteo Amati[3], Luca Gregoratti[3], and Andrei Kolmakov[1*]

[1] NIST, Gaithersburg, MD 20899, USA;
[2] Maryland NanoCenter, University of Maryland, College Park, MD 20742, USA;
[3] Sincrotrone Trieste 34012 Trieste, Italy

*e-mail: andrei.kolmakov@nist.gov

## Abstract

**Additive fabrication of biocompatible 3D structures out of liquid hydrogel solutions has become pivotal technology for tissue engineering, soft robotics, biosensing, drug delivery, etc. Electron and X-ray lithography are well suited to pattern nanoscopic features out of dry polymers, however, the direct additive manufacturing in hydrogel solutions with these powerful tools is hard to implement due to vacuum incompatibility of hydrated samples. In this work, we resolve this principal impediment and demonstrate a technique for in-liquid hydrogel 3D-sculpturing separating high vacuum instrumentation and volatile sample with ultrathin molecularly impermeable membranes transparent to low-energy electrons and soft X-rays. Using either scanning focused electron or synchrotron soft X-ray beams, the principle of the technique, particularities of the in-liquid crosslinking mechanism and factors affecting the ultimate gel feature size are described and validated through the comparison of experiments and simulations. The potential of this technique is demonstrated on a few practical examples such as encapsulation of nanoparticles and live-cell as well as fabrication of mesoscopic 3D-hydrogel structures via modulation of the beam energy**

Hydrogels are a wide class of natural and synthetic hydrophilic porous polymeric scaffolds that can retain a high volume fraction of water, are biocompatible[1] (e.g. poly(ethylene glycol (PEG)-based) and therefore became particularly important for numerous biomedical applications such as extracellular matrix in regenerative medicine, cell transplantations, wounds healing and tissue engineering (see recent reviews[2-5] and references therein). Facile tunability of type and strength of hydrogel chemical bonds, network mesh size, optical, electrical, mechanical and chemical properties is another factor that makes gels highly perspective for biosensing[6], self-healing coatings[7], and soft robotics[8]. In addition, formulation of the composite or hybrid hydrogels[9], where the hydrogel is blended with different types of responsive nanoparticles, polymers or molecules/ions further augments their physicochemical functionalities critical for drugs delivery[10,11], antimicrobial coatings[12], batteries materilas[13], and nanofabrication[14]. Depending on the chemical composition, molecular weight and application of hydrogels, several different triggering agents have been traditionally used for their controlled crosslinking, the major ones being: thermal, photo-, and chemical curing[15]. Modern engineering of 3-dimensional (3D) hydrogel constructs with diffraction limited resolution has immensely benefited from advancements in holographic[16] and additive fabrication approaches using fast photo-induced curing, administered either via diffused or focused laser beam irradiation. Lately, formulation of special photoinitiators for multiphoton laser polymerization being coupled with super-resolution irradiation schemes [17-20] have drastically improved the gel writing resolution to sub 100 nm level as well as the versatility of this method. However, the very low cross-section for multi-photon polymerization and as a result long writing time and potential cytotoxicity of concentrated photoinitiators remain to be impediments for rapid biocompatible 3D hydrogel printing. Alternatively, diffuse γ-, hard X-rays, high energy electron, proton radiations have been widely employed for crosslinking and patterning in the bulk of hydrogel solutions since 1960-s (see reviews [21,22] and references therein). Electron beam lithography, on the other hand, employs a highly focused few nanometers wide electron beam with relatively low energies (1-30) keV and has been successfully used to pattern dry 50-200 nm thick gel films with sub-100 nm accuracy[23-26]. However, high vacuum requirements and therefore solvent-free hydrogels for e-beam writing impede the use of this high-resolution technique for layer-by-layer additive manufacturing. Similarly, the great potential of X-rays for controlled gel polymerization inside live systems[27], particles encapsulations[28] and high aspect ratio structures fabrication[29] have been demonstrated, however, in spite of routinely achievable sub 100 nm resolution in scanning X-ray microscopes in liquid gels[30], no 3D hydrogel patterning in solution with soft X-rays focused beam has been reported yet.

In this work, we introduce a versatile approach to perform focused electron and X-ray beams induced polymerization inside hydrogel solution through ultrathin electron transparent molecularly impermeable membranes separating high vacuum equipment from the vacuum incompatible liquid. Using this method, we were able to perform 3D gel (micro-)printing, cell immobilization and nanoparticle encapsulation inside a liquid solution in a continuous process flow. We define the range of the experimental parameters determining the printing resolution and gel feature sizes and show that the diffusion of radiolitic radicals needs to be invoked to explain the observed systematic differences between model predictions and experimentally obtained feature sizes. We further demonstrate the versatility of the approach by using focused soft X-rays with variable photon energies gel crosslinking with chemical and spatial selectivity and discuss the key similarities and differences in 3D gel patterning mechanism between electrons and soft X-rays.

# Experimental

To deliver focused electron or soft X-ray beams to liquid solutions and for patterning / imaging in liquids we adopted WETSEM[31] (also called liquid cell scanning electron microscopy[32] (LSEM)) methodology and employed custom made (micro-)fluidic or closed chambers equipped with 30 nm to 50 nm thin electron / soft X-ray transparent SiN membranes to isolate the liquid solution from the vacuum of the microscope (Figure 1 (a) for electrons and Figure 1 (b) for X-rays).

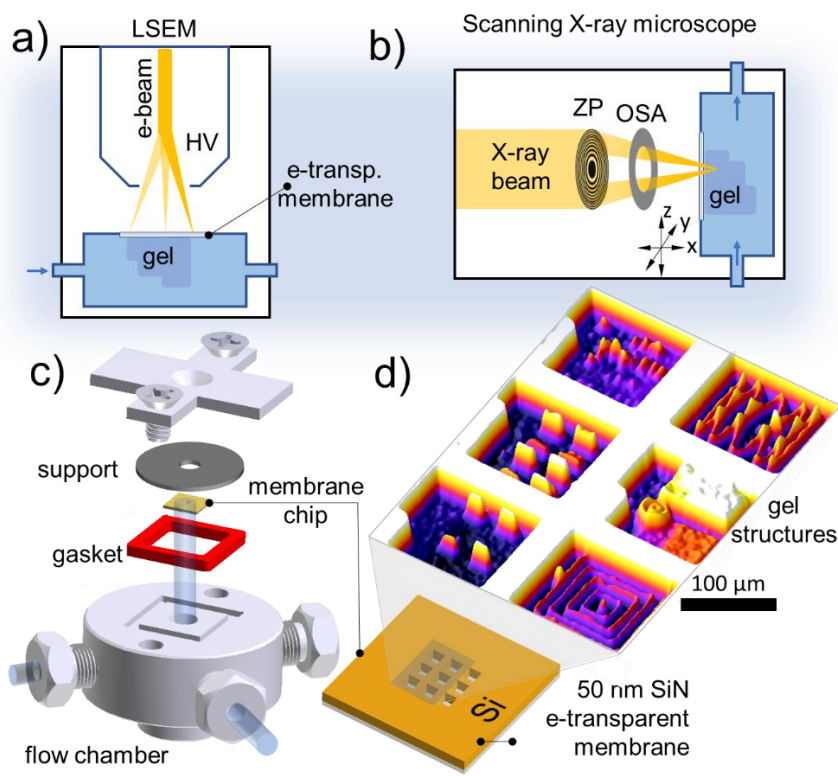

Figure 1. a) Liquid cell SEM (LSEM) setup for site-specific in-liquid hydrogel curing. High vacuum (HV) of the microscope is protected with an electron (e-) transparent membrane. b) Hydrogel curing in the liquid phase using zone plate (ZP) based soft-X ray optics. OSA stands for order sorting aperture; c) the principle design and parts of the liquid sample chamber used in this study; d) 3D-surface plot of an optical image of gel structures electron beam printed on the liquid-facing side of selected six SiN windows

Figures 1c, d demonstrate the experimental setup used in this study for electron focused beam induced curing of liquid gels. The chamber for X-ray studies contained enclosed hydrogel solution and was smaller in size. Both types of chambers were equipped with SiN/Si chip patterned with an array of nine 100 µm x100 µm wide and 50 nm thick SiN suspended membranes. This single-use exchangeable SiN/Si chip is vacuum sealed against the body of the fluidic (or enclosed) cell and the interior of the chamber was filled with Poly(ethylene glycol) Diacrylate (PEGDA) aqueous solution. Nine SiN membrane windows are capable to withstand 1 Bar pressure differential between the liquid sample and high vacuum environments of the microscope and were used for comparative feature writing and combinatorial data collection from multiple windows within a single experiment (Figure 1 (d)). As an example, the arrays of rectangular and fine linear structures were printed on individual membrane windows varying one of the parameters: beam energy, irradiation intensity, step-size size and dwell time at the time. After washing out a solution with water, the dimensions of the crosslinked stable gel structures were then inspected in a hydrated state using AFM and optical profilometry and more precisely in a dry state using SEM. Comparing the height of the same objects in their hydrated and dry state the gel's average vertical swelling ratio was estimated to be 2 ± 0.4 for our gel molecular mass, concertation of the solution and typical electron beam irradiation conditions (see SI Fig S2). The latter calibration was used to evaluate the size of hydrated gel objects based on their SEM inspections in a dry state.

## Working Principle and Major Effects

Upon impinging the liquid interface of the PEGDA aqueous solution, the primary electrons experience a cascade of elastic and inelastic scattering events which slow down and broaden the beam and create a droplet-like highly excited interaction volume where crosslinking of the PEGDA polymer molecules into solid gel takes place (Fig.2(a)). The radiation-induced crosslinking in hydrogel solution occurs via two mechanisms[33]: i) directly, via activating the reactive groups in the polymer with primary or secondary electrons and ii) indirectly (see inset Fig.2a), via electron beam induced water radiolysis that generates a variety of radicals capable to facilitate crosslinking. The partitioning between these two mechanisms depends on multiple parameters such as concertation and molecular weight of the polymer, beam energy, its intensity, exposure dose, solution temperature etc. what was well-studied for highly penetrating ionizing radiation with homogenous excitation and diffusional profiles. We are dealing with an intermediate lesser explored case which involves highly inhomogeneous localized excitation of polymer aqueous solution where beam generated radiolytic radicals can diffuse freely in liquid away from the point of origin thus initiating crosslinking events not only within but potentially beyond the electron interaction volume.

When soft X-rays are used to trigger crosslinking, the net effect is similar, although it proceeds via different interaction pathways. X-ray photon inelastic interactions in a liquid matter are primarily due to photoexcitation of the valence and core electrons of the solute and solution molecules. In the case of water, the relaxation of O1s core hole proceeds primarily via emission of Auger KLL electrons with energies ca 500 eV [34]. After such a de-excitation the energy is deposited to the liquid effectively via the same inelastic electron scattering mechanism as described above for electron beam induced crosslinking. The major differences in electron and soft X-ray induced crosslinking, therefore, are due to their inelastic scattering cross-sections (Fig 2b): i) The values of photoionization cross-section for soft X-rays (100 eV÷2000 eV) are on average ca 100 times smaller compared to few keV electrons and such X-rays, therefore, can penetrate significantly deeper in to hydrogel solution forming gel features with larger heights and lower crosslinking densities (see SI Fig S2c); ii) Electron ionization cross-section of water is a smooth function of energy (Figure 2 b), thus the range of electrons in water solution and printed feature size always increase with energy. In the case of soft X-rays however, the photoionization cross-section sharply increases at the onsets of the specific core level excitations (Figure 2 b). These sudden variations of the X-rays range upon chemical inhomogeneity of the sample will be used to control the aspect ratio of the printed features or chemically selective objects encapsulation.

To gain deeper insight to the factors determining the process of gel printing we compared experimental feature sizes e-beam printed in 20% w/v PEGDA aqueous solution through 50 nm thick SiN membrane with modeled ones applying the same conditions. Monte Carlo (MC) simulated spatial distributions of deposited radiation dose in aqueous solution are depicted in Figure 2c for a few different energies and 5 nm wide electron beam. Note: the gradual change of the absorbed dose also implies the significant variation of the crosslinking density (and therefore mechanical and other gel properties) across the thickness of the printed object. Assuming the critical energy dose for gelling being in excess of ca$10^3$ Gy[33], the expected height of stable hydrated objects ranges from ca 200 nm for 3 kV to ca 8 microns for 20 kV electron beams (Figure 2 c).

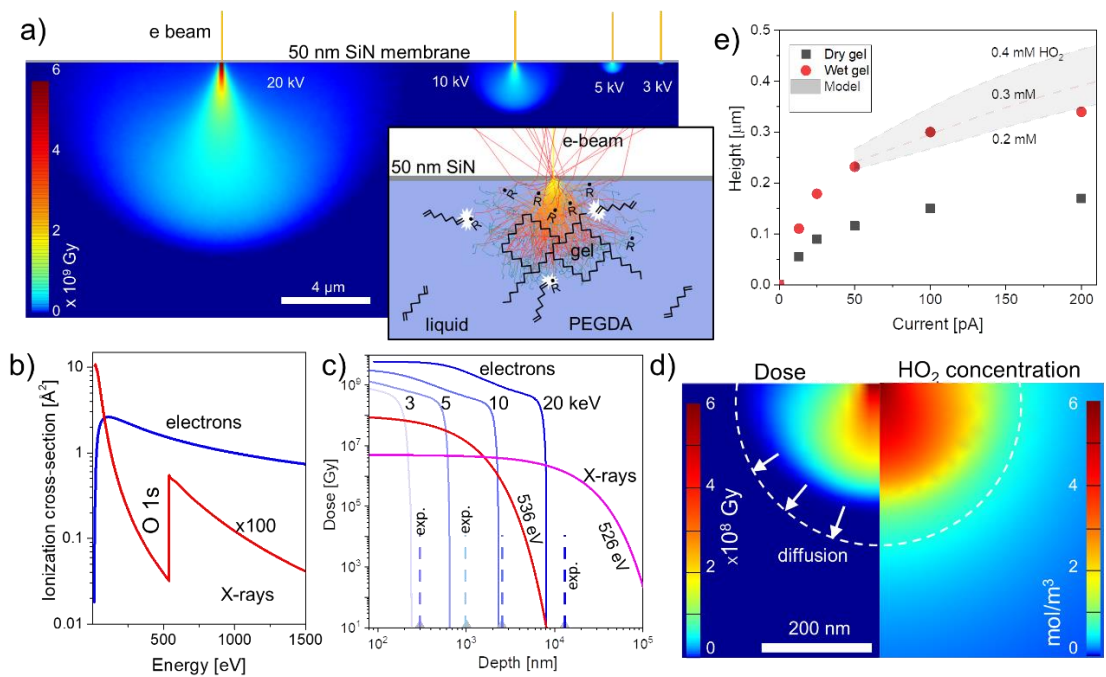

Figure 2. (a) Spatial distribution of the energy deposition by 5 nm wide 20, 10, 5, and 3 kV (left to right) electron beams in water calculated using MC simulations. Inset depicts the concept of direct and indirect (through radiolytic radicals *R*) PEGDA polymer crosslinking (b) the comparison of the electron and soft X-ray ionization cross sections at low energies; (c) MC simulated electron and X-ray energy doses as a function of depth for electrons (100 pA current, 1 ms dwell time) and for X-rays ($2 \cdot 10^9$ photons/s and 10 ms dwell time). These parameters represent general settings used in this work for feature printing. Dashed vertical lines depict experimentally observed hydrated feature sizes created using the same exposure parameters. Lines color coding is the same as for simulated curves with corresponding electron beam energy. d) Energy deposited distribution for 3 kV 200 pA electron beam (left panel) and the corresponding distribution of hydroperoxyl radical concentration *(right panel)* calculated using beam induced radiolysis reaction-diffusion model. The scale size is the same for left and right panels and highlights the effect of diffusion of radiolytic $HO_2$ outside the region where the energy is deposited; The deviation of the size of the printed objects in hydrated state (arrows and dotted curve) from the one defined by electron range manifests the effect of diffusion of radiolitic species on a feature size. (e) Heights of the dry gel features formed at 3 keV as a function of electron beam current *(grey)*, and their estimated values *(black)* in a wet state based on AFM calibrations. The shadowed band represents the estimated feature height based on MC simulations considering $HO_2$ radical as crosslinking initiator with a concentration between 0.2 mM and 0.4 mM.

However, the experimental data on thicknesses of hydrated features obtained under the same conditions (dashed lines in the Figure 2c and Figures S3, S4 SI) are noticeably and systematically larger than electron ranges obtained from Monte-Carlo simulations, what implies that indirect crosslinking via runaway diffusion of radiolitic activators contributes to size increase of the printed features. To evaluate the latter more accurately we adapted a kinetic radiolysis model, previously developed for environmental SEM [35] and in-liquid transmission electron microscopy[36,37], to our SEM conditions (see details in SI). The numerical simulations based on this model predict short lifetime and therefore a small diffusion length (< 200 nm) for most abundant crosslinking initiator OH radical outside the electron beam interaction volume, thus ruling out it from being a primary height determining factor. Other possible radiolytic crosslinking activators e.g. hydroperoxyl radical, have significantly larger lifetime and can, therefore, diffuse to longer distances (1-5 microns) before reacting out completely. Figure 2 d compares the energy dose distribution upon water irradiation with 3 keV focused beam (left panel) with corresponding $HO_2$ concentration profile (right panel). As can be seen, $HO_2$ crosslinking activator concentration remains appreciably well beyond the electron range and therefore may account for the systematically increased size of the printed features. For practical applications, it is useful to estimate the critical concentration of radiolytic activator required to crosslink the PEGDA hydrogel. Comparing the experimental feature thickness in a wet state and the range of modeled concentration profiles of hydroperoxyl radical for 3 kV, we evaluate a critical concentration for crosslinking of the hydrogel at the given concentration and molecular weight of PEGDA solute to be ca 0.3 mM of $HO_2$ (Fig.2e).

## Printing Controls

We now discuss the primary experimental parameters that can be tuned to control the size and shape of the printed objects. In the raster scanning mode these are: i) electron beam energy ($E$), ii) Dwell Time ($\tau_D$) at a pixel, iii) pixel size (also step-size) during the scan ($L$) and iv) exposure dose $D$ per pixel defined as $D = I_B \tau_D n / L^2$ where $I_B$ is an electron (photon) beam current (intensity) and $n$ is a number of scans.

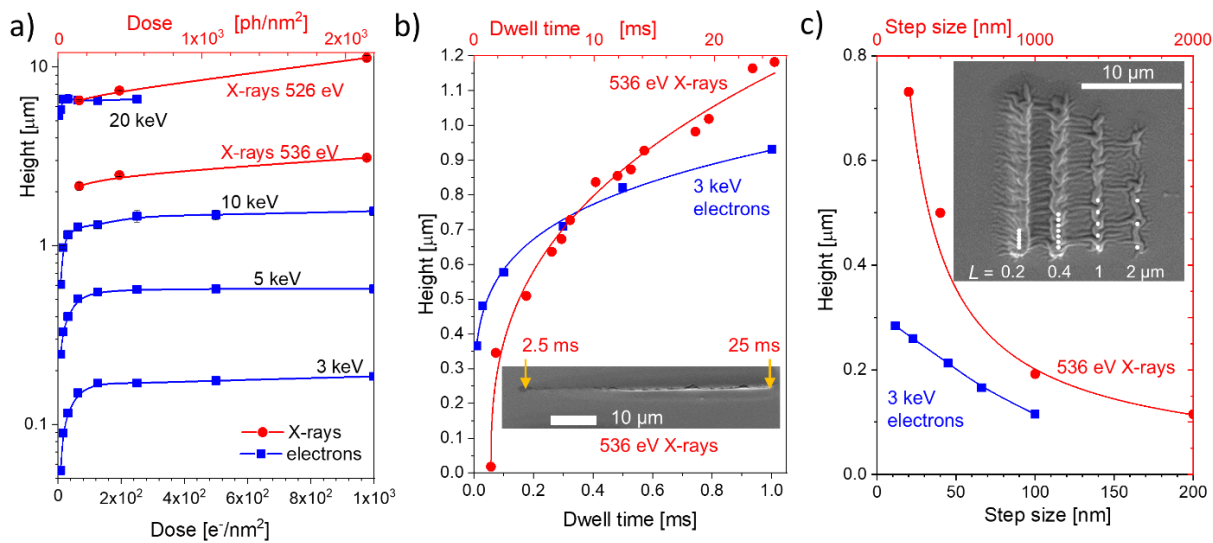

Figure 3 SEM measured heights of the dry rectangular and linear gel features as a function of a) exposure dose (via beam intensity increase and fixed dwell time) for different energies of electrons (blue) and X-rays (red) beams. The anticipated drastic reduction of the feature size created with photons with post O1s edge energies can also be observed; b) dwell time for 5 keV electrons and 536 eV X-rays (beam intensity fixed). An inset shows SEM image of gel structure written with variable along its length using 536 eV X-rays; c) step-size for 3 keV for electrons (blue) and 536 eV X-rays (red). Inset shows SEM image of gel lines written with different step size using 536 eV X-rays

Figure 3 shows the experimentally measured heights of the crosslinked gel dried features as a function of each of these parameters with other being fixed. In addition to the apparent increase of the feature size with electron and X-rays beam energy, the height variation with exposure dose has characteristic fast rise followed by saturation behavior (Fig. 3a) also commonly observed for dry films. In accordance with prior reports, the certain dose threshold (ca 4 e$^-$/nm$^2$ in our case) is required for through-membrane crosslinking of a stable gel structure. While the increase of the dose does not affect the dimensions of the interaction volume in water, it expands the boundary at which the critical concentration of hydroperoxyl radical can be reached. As a result, feature sizes increase with the current until the point when radical concertation saturates at steady state values being consumed by other species. The same is valid for the feature heights increase with the dwell-time (Figure 3 b). *Step-size* becomes a rather important parameter when the beam is rastered across the sample surface. By increasing the step-size and therefore the pixel area, one can tune the overlap between interaction volumes and diffusion zones of individual adjacent pixels, thus reducing or increasing the effective thickness (and width) of the printed feature. An important distinction between the electron and soft X-ray beam writing used in this study is the size of the probe: electron beam has a diameter of ca 5 nm while for X-rays it is ca 150 nm. On the other hand, the effective diameter of the interaction volume of a few keV electron beam is appreciably larger (see Fig 2a) compared to one for soft X-rays. Therefore, for X-rays, the formation of corrugated/ discontinuous patterns can be observed as soon as the step-size become larger than 150 nm (see inset in Figure 3c). On the contrary, few keV electron beams generate continuous patterns, for step-size values even larger than 100 nm.

To summarize: for practical applications, *Dwell-time* and *Step-size* can be useful independent parameters to control the size and crosslink density of printed gel structures. Unlike the *Beam Energy* and *Intensity*, these two parameters are untangled in SEM can be tuned during the scanning without the need for re-focusing of the microscope. In the case of the e-beam printing, the lateral and longitudinal resolution of the features obtained are coupled and are proportional to each other (Fig.2a). As one intensifies any of the parameters: *Beam Energy, Intensity, Dwell time*, both the width and height of the pattern increase. Finally, the smallest feature size and maximum attainable resolution for SEM based gelation are dependent on the minimal energy of electrons that can penetrate through the SiN window and the threshold dose required to crosslink the polymer in solution. We were able to routinely write ca 150 nm thin and 100 nm wide gel lines with through a 50 nm SiN window using 3 kV electron beam energy, 13 pA, 1 ms dwell time and 100 nm step-size (see S1 of SI). Sub-100 nm features are attainable if thinner membranes or membrane-free printing scheme (e.g. using ASEM) be used.

# Application Examples

### *Composite hydrogels*

The applications of hydrogels became immensely broadened via synthesis of composite formulations, which allows for the engineering of their optical, electrical, mechanical, magnetic properties for numerous applications. Fabrication of composite hydrogels can be broadly classified into two methods: (i) *in-situ* ones where functional inclusions (e.g. nanoparticles) or precursors are premixed in the polymer solution and become stabilized in the hydrogel during the crosslinking process, (ii) whereas *ex-situ* techniques typically involve an inclusions impregnation process, applied after the crosslinking of the host matrix. *In-situ* encapsulation offers advantages of embedding guest objects independently of their size, homogeneously across the bulk, while the post-crosslinking impregnation depends on surface-to-bulk diffusion of chemicals in the gel's matrix which is effectively hampered for the objects larger than gel's mesh size. Figure 4 a (panels 1 and 2) show the principle of in-liquid entrapping of nanoparticles via the use of the focused e-beam induced cross-linking of nanoparticles suspension followed with SEM characterization of the composite gel. Monte Carlo simulations of electron beam energy deposed into the gel colloid (Figure 4 b) predict an effective immobilization of high Z nanoparticles in the hydrogel matrix due to the formation of gel cocoon around such a nanoparticle via preferential crosslinking of the near particle polymer solution by secondary and backscattered electrons. For many applications, it is important to image the encapsulated particles inside the gel with high spatial resolution. Figure 4 c depicts the SEM images of mixed 75 nm Au and Ag nanoparticles entrapped inside the crosslinked hydrogel cube recorded through 50 nm SiN membrane in a hydrated (inset) and dry state (background image) using backscattered electrons (BSE) detector sensitive to fast electrons. In this SEM imaging mode, the contrast of the objects is determined by the difference of the atomic numbers (Z) of the nanoparticles and matrix material as well as on the depth at which the electrons are collected. In the SEM image Au and Ag particles with much larger Z compared to hydrogel matrix appear brighter, and both: their signal strength and the resolution wanes with the depth of the nanoparticle inside the gel (Figure 4 c). Both images show similar and homogeneous immobilized particles distributions and energy-dispersive X-ray spectroscopy (EDS) chemical maps discriminate between Au and Ag nanoparticles. Overall, SEM can be used to probe gel encapsulated high Z nanoparticles as deep as few hundred nanometers using a 30kV beam with resolution still better compared to conventional optical microscopy.

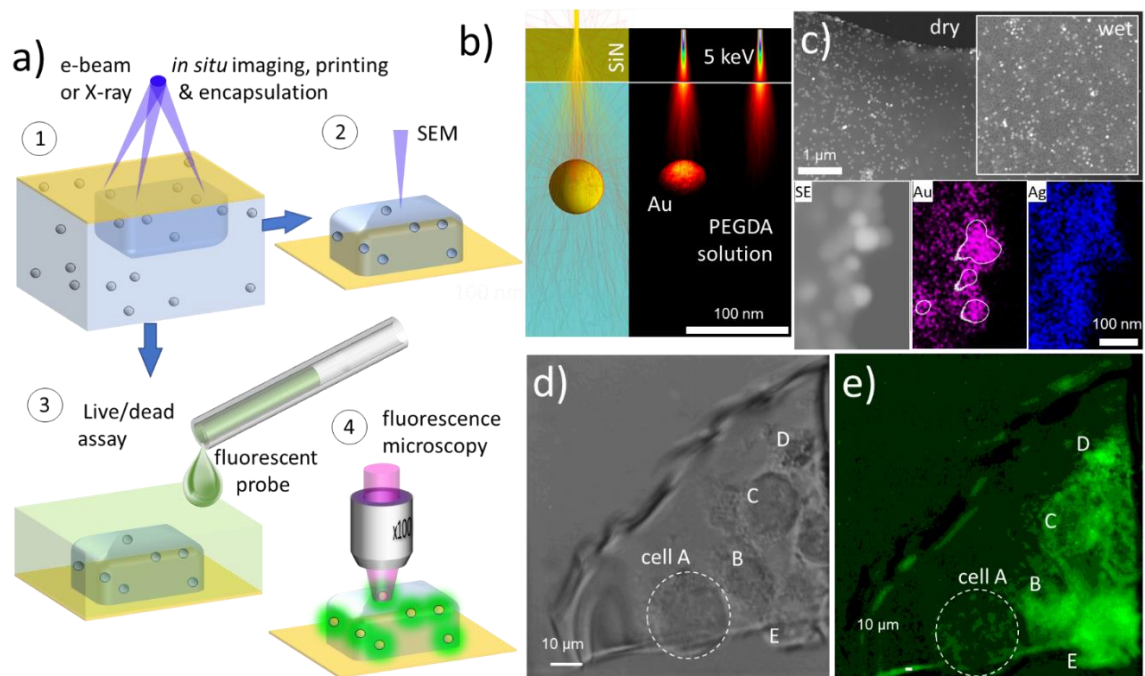

Figure 4 a) The principle of e-beam induced encapsulation and characterization of objects inside a composite gel. (1) Irradiation and patterning of the gel solution through the SiN membrane, followed with SEM characterization in a dry stage (2). (3) Hydrogel encapsulated cell treated with a fluorescent probe for viability tests. (4) Fluorescence microscopy of biological objects encapsulated in a gel. b) Monte-Carlo simulations of the electron trajectories for Au nanoparticle immersed in hydrogel (left panel) and corresponding deposited energy distribution inside the particle and nearby liquid (right panel) upon excitation with 5 keV electron beam. c) SEM image of embedded 75 nm Au and Ag nanoparticles collected in a dry (left panel) and hydrated state (right panel). Bottom raw depicts the zoomed area and corresponding Au and Ag chemical maps collected with EDS. d) Optical image of the hydrogel gel after electron beam irradiation of cell-laden PEGDA solution. f) Fluorescence microscopy image of gel embedded cells stained with viability indicator (calcein green dye).

*Cells immobilization*

Though the viability of cells in PEG-based hydrogels is well established[1], most of the previous encapsulation studies have been performed using photo-initiated crosslinking[38]. Here we attempt to employ *in-liquid* gel focused electron (X-ray) beam crosslinking technique to encapsulate live cells before their lethal radiation threshold dose is reached. The extent of radiation damage of the biological objects (cells) during electron beam-induced PEGDA crosslinking is not well known and generated radiolytic species like $OH^*$, $O^-$, and $H_2O_2$ at high concentrations can be detrimental to the cells. Prior electron microscopy studies of biological specimens in the wet environment reported a broad range of critical dose values from ca $10^{-3}$ $e^-/nm^2$ to ca $10^2$ $e^-/nm^2$ that is considered acceptable for live mammalian cells, yeasts and other microorganisms. This ambiguity manifests a fundamental challenge of high-resolution electron microscopy of live cells as well as the variance in live/dead criteria applied (see refs.[39-43] as an example and discussion therein). As we showed above, PEGDA crosslinking threshold dose is ca 4 $e^-/nm^2$ in our setup and can be further reduced if higher molecular weight PEG is used.[23] Since the imaging during crosslinking is not a requirement, the encapsulation of live microorganisms using our approach can be feasible. The test process flow is depicted in Figure 4a, panels 1, 3, 4. Once a polymer solution with premixed live cells was exposed to electrons, the crosslinked gel with encapsulated cells was tested with standard calcein-AM cell viability assay[44]. In this test, live cells uptake the non-fluorescent calcein-AM ester. Inside the living cell, ester reacts with cytosolic esterases which convert it into green-fluorochrome: calcein to which cellular membrane is not permeable. Bright-field optical image in Figure 4 d shows SiN window containing cell-laden PEGDA solution after exposure to 10 kV primary beam with an average exposure dose of 8 $e^-/nm^2$ (absorbed dose ca 3 x $10^6$ Gy). Fluorescent microscopy image of gel-immobilized cells in the Figure 4f complements Figure 4e and indicates (i) that cells B-E do produce fluorescent calcein after encapsulation; (ii) calcein distribution remains confined within the cellular borders and (iii) some of the cells (cell A in the figure 4e) appear dark implying its necrosis, while the rest of the cells presumably survive the encapsulation procedure. It is necessary to note that the local absorption dose at the point of beam incidence is a few orders of magnitude higher than the cells lethal dose limit. However, as can be seen from Figure 2a, c, d, the radiation damage is localized within a micron wide region near the SiN membrane. Thus, the cells floating in solution a micron or more away from the membrane see a significantly diluted load of radiolytic species. In addition, the fraction of the radiolytic species produced by the beam becomes scavenged during gel crosslinking reactions what, therefore effectively reduce the concentration of toxic species seen by the cells. This is a promising result which, however, has to be considered as preliminary observation and requires further studies.

   *3D printing*

The panels in Figure 5 show exemplary 3D structures printed using *in-liquid* crosslinking approach and the capabilities and limitations of the technique for both electron (Figure 5(a-f)) and soft X-ray (Figure 5 (g-i)) focused beams are highlighted below. The technique benefits of high cross section excitation process, was well-tested on standard dry gel films lithography and can complement the existing state of the art optical 2PP crosslinking methods with faster writing time and potentially higher resolution. The ultimate resolution is restricted by the mean free path of primary and secondary electrons in a liquid which can be as short as ca one nanometer for 50 eV÷80 eV electrons[45]. On the other hand, through-the-membrane electron transparency dictates the minimal energy required for crosslinking, thus nanometer size resolution can only be achieved with graphene-like ultrathin membranes. The primary energy of electrons also determines the range of electrons in water and therefore the upper limit of the structure's heights to ca 10 micrometers when standard 30 keV SEM is used (Fig. 5 b, d). The lateral dimensions of the printed object are restricted by a few hundred microns depending on the mechanical stability of pressurized 30-50 nm thick SiN (or $SiO_2$) electron transparent membrane. In addition, through-membrane approach implies limitations on direct writing of suspended or narrower footprint structures. However, the aforementioned limitations are not finite and arbitrary shaped millimeters or even centimeters size features with tens nanometers resolution can in principle be fabricated via curing the open liquid surface directly using atmospheric SEM chamber-less setup[46].

The coaxial cylindrical 3D structures in the Figure 5 d, e have been printed in liquid gel solution via varying only one electron beam parameter: electron energy in (5 d) or dwell time in (5 e). The dynamic range of heights exceeding 100 was achieved via beam energy variation, however, the sharpness of the features drops concomitantly with energy. The facile and instant modulation of the irradiation parameters such as dwell time and step-size allow batch-fabrication of the high aspect ratio microstructures as a flagella-like object in Figure 5 c in a single run within a fraction of second. The similar structures have been used for locomotion at low Reynolds numbers after being functionalized with magnetic nanoparticles[47].

The effect of the beam intensity and writing sequence on the linear feature size and morphology is shown in Figure 5f. As discussed above, the feature size rapidly increases with beam intensity and then saturates. Interestingly, the nodes height increase occurs for thinnest overlapping lines and is negligible for thick nodes. This has a direct consequence for additive fabrication of overlying structures and indicates that crosslinking still proceeds on top of already printed structure under the conditions where either (i) electron range (together with radiolitic initiator diffusion length) exceeds the size (height) of the first printed layer or (ii) the saturation of the crosslinking within the interaction volume has not been achieved during the first layer printing.

3D printing with focused variable energy X-ray beam has an additional capability to modulate the feature size via printing with different photon energies just below and after the element specific absorption edge. The examples in the Figures 5 h, i show that it can be performed in both ways: through already printed feature as in the Figure 5h and as a separate nearby structure as in the Figure 5i. Compared to printing with electrons, the height of the X-ray induced structures was appreciably larger while their density correspondingly lower what results in the significant surface rippling of the features upon drying.

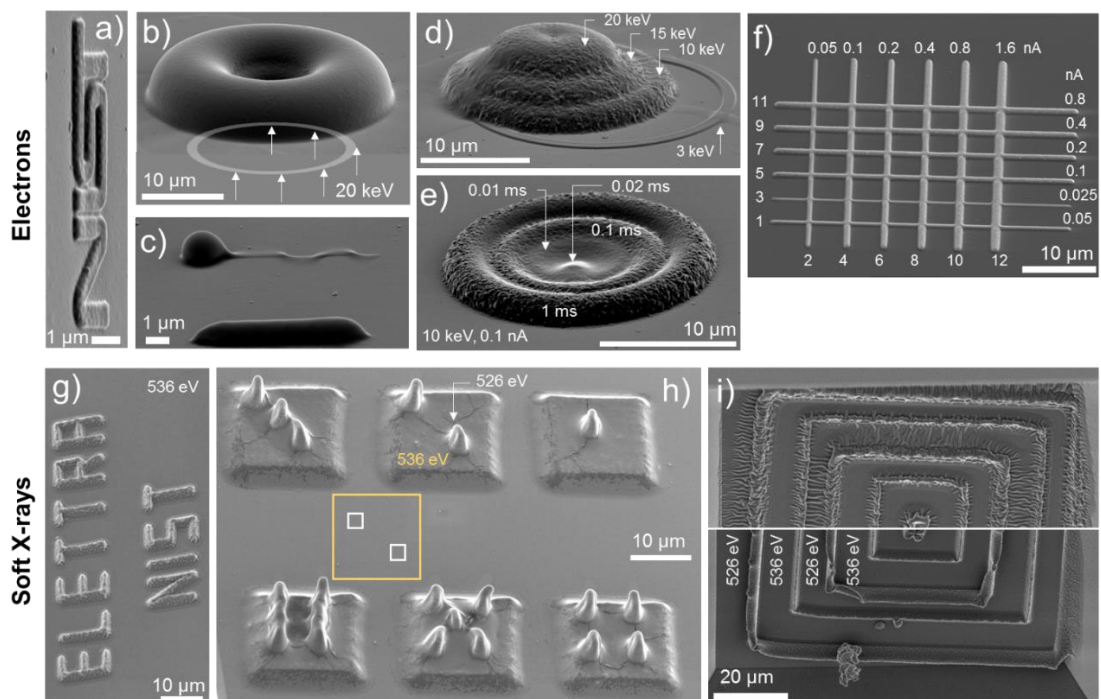

Figure 5 SEM images of exemplary 3D features fabricated using focused electron beam a) - f) and soft X-rays g)- h). a) NIST logo is written with 3 keV electrons (100 pA 1msec dwell time) using 10 nm pixel size. b) Donut-like gel feature made by writing 1 µm wide ring with a radius of 10 µm using 20 keV electrons 100 pA, 1 msec dwell time. c) Flagella-like structures printed with 10 keV, 3 pA, 10 µsec dwell time and 100 nm step size. d) Dome structures formed out of four overlapping coaxial rings by varying electron beam energy for every ring; Parameters: beam current 100 pA, 50 nm step-size and 1ms dwell time. e) Similar to (d) concentric rings formed at 10 kV by varying dwell time and pixel size. f) Grid structure formed by varying beam currents at every line. Electron beam: 5 kV, 0.01 msec dwell time and 100 scans per line. g) ELETTRA and NIST logos printed with 536 eV 150 nm wide X-ray beam with 25 msec dwell time. Photon flux ca $2\cdot10^7$ ph/nm² s and 100 nm step size. h) Dices printed with two photon energies: 13 µm x 13 µm base squares- 526 eV, 2.5ms dwell time, flux $1.3\cdot10^5$ Ph/nm²s and small 2.5 µm x 2.5 µm squares 536 eV, 2.5 msec dwell time, 100 nm step-size and $1.2\cdot10^5$ ph/nm²s flux. i) Labyrinth structures with walls printed either with 526 eV or 536 eV, 100 nm step-size and 5 msec (top panel) and 25 msec (bottom panel) dwell times.

# Conclusions

We employ scanning electron and X-ray microscopy for spatially controlled cross-linking of hydrogels in their natural liquid state. The feasibility of this technique was demonstrated via gel encapsulation of live cells, fabrication of composite hydrogels and 3D printing of model hydrogel structures with submicron resolution. We evaluated the threshold dose required for electron beam induced cross-linking in a liquid state and explored the effect of diffusion of radiolitic species and other experimentally tunable parameters such as electrons (or X-rays) energy, beams intensity, exposure time etc. on the resolution and size of the features formed. High spatial resolution printing of a large class of hydrogels in the liquid state can also be extended to gas phase polymerization and offers unique advantages in shape, size and precision compared to traditional dry gel lithography and significant improvement in writing time compared to multiphoton polymerization methods. The proposed technology can be implemented in any high vacuum, environmental or atmospheric pressure (or air) SEM (ASEM) or laboratory-based X-ray microscopes. The ability to operate with free liquid replenishable surfaces offered by ASEM is particularly attractive since it allows for truly additive nanofabrication and rapid prototyping of gels with sub-100-nm resolution using this method. The tunability of X-ray energy at synchrotrons opens an additional opportunity to conduct element specific 3D gel printing in solutions relevant to biomedical, soft micro-robotics, electrochemical and other applications. Moreover, the combination of our method with recently proposed implosive fabrication technique[14] can in principle result in nanometer-scale 3D printing.

# Methods

## Hydrogel Synthesis and Printing

The standard tests have been performed with a stagnant liquid setup where ca 10 mL of 20% w/v PEGDA (average molecular weight 0.7 kg/mol) solution was drop-casted on to Si chip with nine 50 nm thick SiN membrane. The chip was connected sealed with vacuum-tight chamber. The assembly was placed inside standard SEM where the liquid polymer solution was patterned with a scanning focused beam through the SiN window with a known amount of dose at every pixel. The experiment was repeated in another part of the window or at the different SiN window using a different set of irradiation parameters. After e-beam exposure, the chip was dismounted from the chamber and rinsed in water to remove the unreacted solution. This leaves an array of printed gel features on SiN membranes which were inspected in a hydrated state with AFM or optical microscopy/profilometry and/or in a dry state using SEM, AFM, EDX, XPS, µ-Raman and other characterization tools. Printing with soft X-rays was performed at ESCA microscopy beamline at ELETTRA equipped with zone plate optics capable to focus monochromatized light to a spot 150 nm-200 nm in diameter. The undulator and monochromator have been set to operate either at 530 eV or at 536 eV with the photon flux in the order of $10^9$ ph/s at 150 nm wide focal spot. The chamber equipped with the same chip with nine SiN membranes array was filled with PEGDA solution, sealed and scanned in front of the beam in a pre-programmed path to generate a required pattern.

## Composite Hydrogels

Gold and silver nanoparticles (75 nm in dia.), mixture (ca 1:10 wt/wt) suspension in water was pre-concentrated by centrifuging (2000 r/min, 5 min) and was subsequently extracted and mixed with the 20% w/v PEGDA solution. As prepared composite solution was irradiated in the liquid phase with an electron or soft X-ray beams through 50 nm SiN membrane. The chip with the printed composite gel structures was then developed in water and subsequently analyzed as described in the article and Supporting material.

## Cell Encapsulation and Proliferation Tests

Caco-2 cells were thawed and cultured in DMEM (Dulbecco's Modification of Eagle's medium) with 4.5 g/L glucose and L-glutamine without sodium pyruvate for a few days. These were subsequently washed in PBS and DMEM. Lifting off process was carried out using 2 ml of .05% (w/v) Trypsin-EDTA and left for 5-10 min until the attached cells become mobile on the slide. Neutralization of Trypsin is done by adding an equal volume of growth medium. The obtained cell suspension is concentrated by centrifuge. The cell concentrate is added to the PBS based PEDGA solution and cells were allowed to adhere to the SiN membrane. For viability tests after irradiation, the crosslinked gel with encapsulated cells was rinsed in the growth medium for cell and exposed to calcein green dye in the growth medium for 1 hr. Inside the live cells, the non-fluorescent calcein is converted into green-fluorescent calcein via de-esterification of the acetoxymethyl group by the esterases only produced by a live cell.

## Modeling Details

A stack of 50 nm SiN and 20-micron thick water layer was modeled with the electron beam incident on the SiN membrane. The Monte-Carlo (MC) simulations (described in the SI section) generated the trajectories and corresponding energy deposited (Gy). The parameters used to Generate Figure 2 a) were as follows: 625,000 electrons for a 5 nm beam diameter for 3 keV, 5 keV, 10 keV, and 20 keV primary beam energy. The energy deposition results in Gy for 625000 electrons were scaled depending on the current value to obtain the rate of energy deposition (Gy/sec) and fed into the radiolysis kinetics model (described in detail in the SI section). To generate results shown in Figure 2 d), the CFD model was executed for 3 kV primary beam, and currents current values: 50, 85, 125, 160, 200 and 215 pA.

## Acknowledgments:

TG acknowledges support under the Cooperative Research Agreement between the University of Maryland and the National Institute of Standards and Technology Center for Nanoscale Science and Technology, Award 70NANB14H209, through the University of Maryland. Authors are thankful to Dr. Andras Vladar, Dr. John Villarrubia, Dr. Dean Delongchamp (all at NIST), Prof. G. Kolmakov (CUNY), Prof. Dr. Michael Zharnikov (Univ. of Heidelberg) for constructive feedback on the manuscript and to Dr. D. Perez (NIST) for the help with profilometry measurements.

Disclaimer: Certain commercial equipment, instruments, or materials are identified in this paper to foster understanding. Such identification does not imply recommendation or endorsement by the National Institute of Standards and Technology, nor does it imply that the materials or equipment identified are necessarily the best available for the purpose

## Author contributions

A.K. and T.G. conceived the idea and conducted core measurements. A.K. directed the work. T.G. performed experiments, data analysis, simulations, and modeling. G.H. provided engineering support to the method. E.S. performed AFM measurements. J.S. did electron beam lithography. Y.Y. and M.E. grew and characterized cell cultures. V.A. and T.G. conducted profilometry characterizations. X-ray studies have been performed by P.Z., M.A., L.G., and A.K. T.G. and A.K. wrote the paper, and all authors commented on the manuscript.

## Additional information

Additional information Supplementary information is available in the online version of the paper. Reprints and permissions information is available online at www.nature.com/reprints. Correspondence and requests for materials should be addressed to A.K.: andrei.kolmakov@nist.gov

## Competing financial interests

The authors declare no competing financial interests.

## Supplemental Material

**Height estimation**

The dimensions of the obtained features were measured using various techniques, including optical Profilometry (Figure S1 a, d), SEM (Figure S1 b, e, f) and AFM (Figure S1 c). Depending on the shape and size of the features different methods were found suitable.

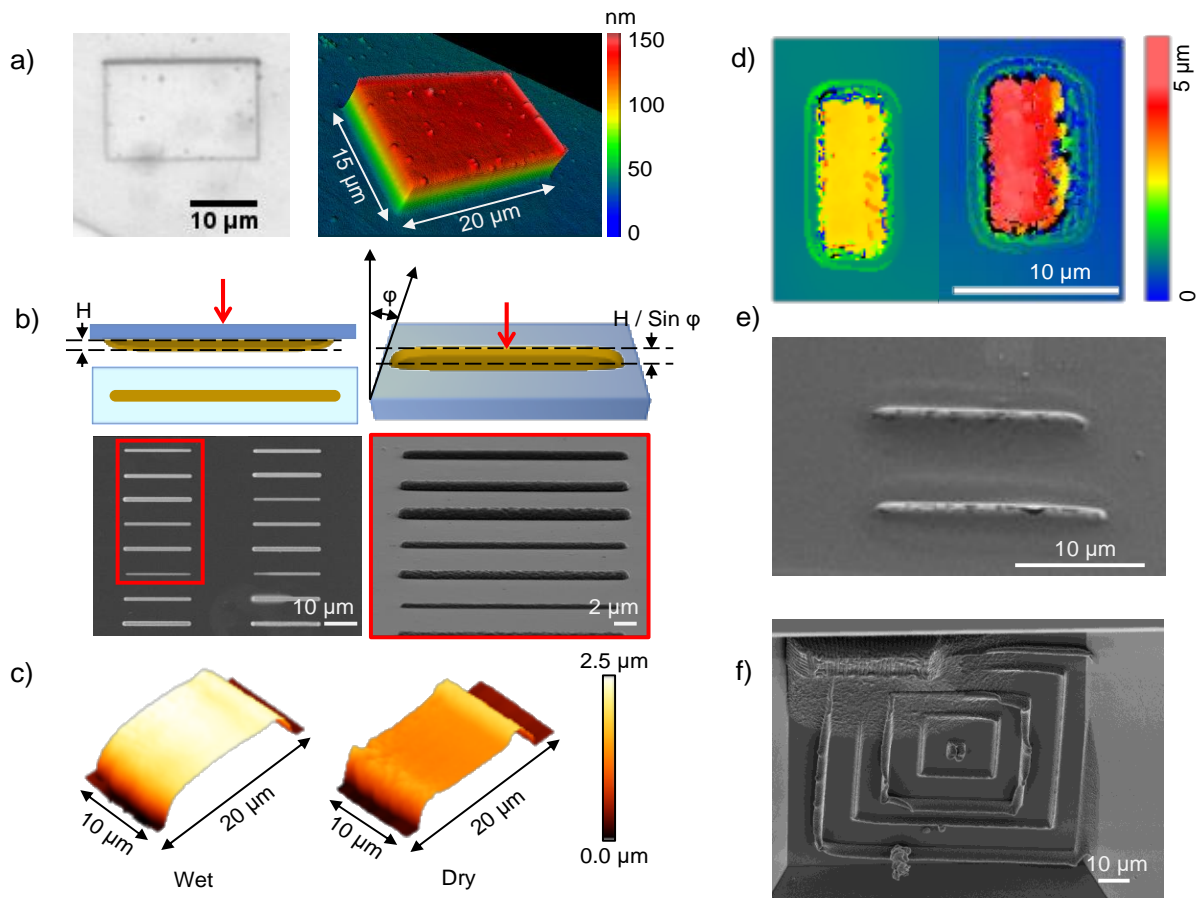

Figure S1 Different methods of height estimation are shown for a), b), c) electron crosslinked samples and d), e), f) X-ray crosslinked samples. a) Profilometry: (left) optical image, (right) 3D structure of the hydrogel formed at 3 kV. b) SEM: (left) image of thin lines taken through the membrane; (right) the same sample was placed upside down and tilted to view projection of the cross-section. Height was estimated based on the angle of tilt. c) AFM of hydrogel done in a hydrated state (left) and same feature done in a dry environment (right). d) Profilometry of 536 eV X-ray sample (left) and 526 eV X-ray sample (right). e) SEM of X-ray lines formed at 526 eV. f) SEM of the labyrinth-like structure formed by squares alternating from 526 eV to 536 eV from the center out.

Samples with larger dimensions were quantified using profilometry. For laterally thin samples, SEM was found to be most suitable. The sample was mounted on a tilted stage facing the electron beam. The projected height of the sample can be measured and used to calculate the actual height based on the tilt angle as shown for electron samples in Figure S1 b. For vertically thin features generated using low electron beam energy, AFM (Figure S1 c) was found to be most accurate for height estimation. Similar height estimations were done for X-ray crosslinked samples as shown in Figure S1 d, e, f. For consistency, all measurements were done after exposing the sample to vacuum.

Since these hydrogels have a high-volume fraction of water, they shrink when exposed to air or vacuum. This is deduced by measuring the shrinkage of macro-sized UV cured hydrogels with time using an optical microscope, as shown in Figure S2 a. UV cured samples shrink to 20% of their original wet size. However, there can be differences in the water content of the UV cured samples and the electron beam cured samples depending on the density of crosslinking. In order to estimate the original dimension of electron beam cured wet hydrogel, AFM was done in the hydrated state in a liquid environment and then post drying on the same feature. Results, shown in Figure S2 b, suggest an average shrinkage of 50% ± 20% on vacuum drying. This is significantly less when compared to shrinkage fraction in UV cured samples (80%) suggesting differences in the curing mechanism of the two techniques. The difference between these two techniques can be seen in the SEM images in Figure S2 c, where the UV cured samples exhibits rougher morphology upon drying indicative of a more porous structure. For comparison SEM image of X-ray crosslinked sample is also shown, reflecting patterns which are intermediate in size, between the UV and electron crosslinked samples. We conclude that the original dimensions of the as-prepared hydrated hydrogel via electron beam and X-ray samples are therefore approximately twice and two-four times as large compared to the dry values, respectively. All dimensions shown in the main text Figure 2 are from dry samples.

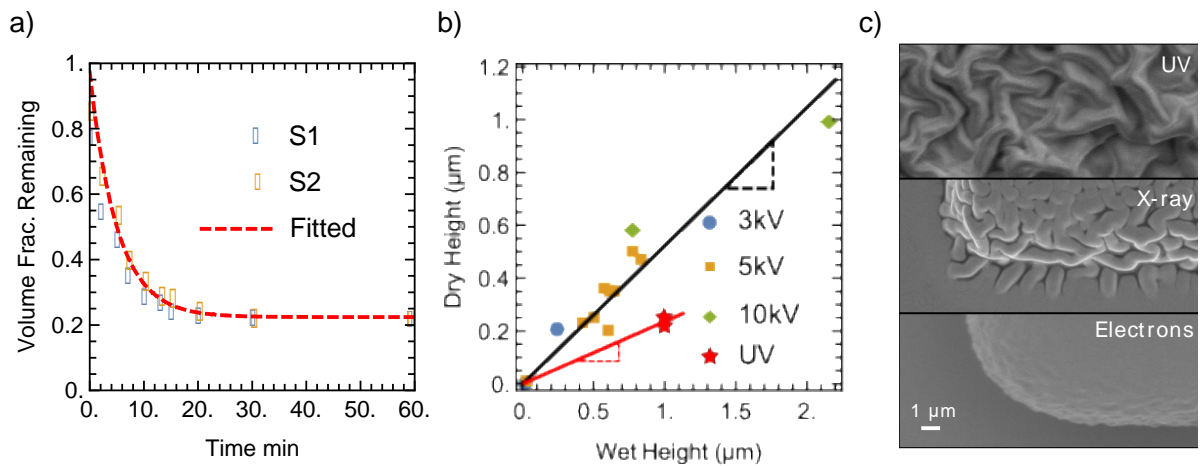

Figure S1 Effect of drying on UV cured and E-beam cured samples. a) Volume Fraction remaining of UV cured samples S1 and S2 as a function of time b) dry height vs. wet height of different samples to estimate the dry fraction. Average Dry fraction for e-beam samples 0.52 (black line) and for UV cured samples is 0.2 (red line). C) SEM image of UV cured, X-ray cured and electron beam cured samples at similar settings and magnification after drying.

A correction factor of 2 is therefore multiplied to the dry height of the electron beam samples to obtain the wet height.

**X-ray Dose Estimation**

Intensity attenuation of X-rays with depth can be calculated using Beer-lamberts law

$$N = N_0 \exp(-\mu d) \quad (1)$$

here $N_0$ is the photon flux per unit area at the surface, $\frac{1}{\mu}$ is the attenuation length and $d$ is the depth. Dose per unit mass can then be calculated as

$$-\frac{1}{\rho}\frac{\partial (N\,h\nu)}{\partial d} = \frac{\mu N_0 h\nu}{\rho} \exp(-\mu d) \quad (2)$$

here $\rho$ is the density of the interacting media and $h\nu$ is the photon energy. Beam shape at the sample was known to be Gaussian with FWHM of 150 nm. Measurements from photodiodes were calibrated to obtain the total photon flux on the sample. From the beam shape and total flux, $N_0$ was calculated by averaging over the pixel area (same as step size).

**Electron Dose Estimation**

The spatial distribution of energy is computed using Monte-Carlo simulations[48], where a primary beam of known energy is allowed to interact with a stack of 50 nm Silicon Nitride membrane and bulk water. The electrons are allowed to experience elastic and inelastic collisions in a cascade-like process as they travel until they reach threshold energy and thermalize. The elastic interactions were treated as discrete events using Motts cross-section, whereas the inelastic events were approximated based on the mean energy loss model by Joy & Luo [49]. Figure S3a shows the trajectory of electrons with the color denoting the energy of electrons for 5 keV primary beam. The corresponding energy deposited distribution into the water is shown in Figure S3 b.

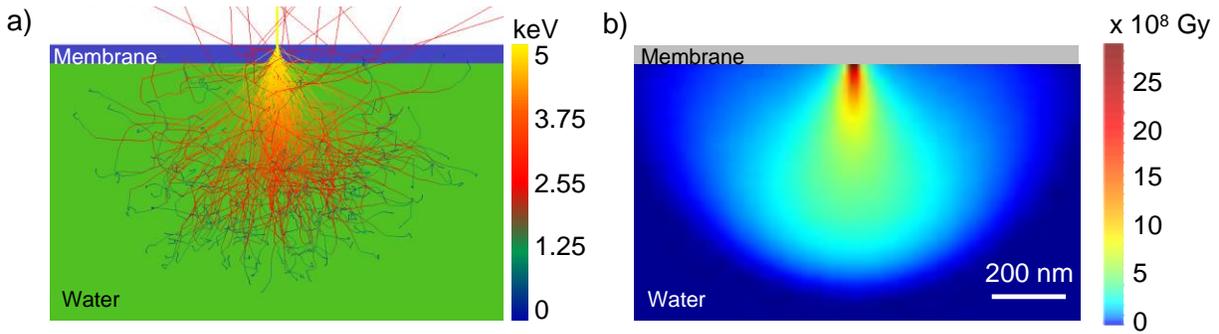

Figure S3 Results from Monte-Carlo simulation generated by simulating 625000 electrons, for a 5 nm beam diameter at 5 keV primary beam energy. a) The trajectory of electrons, with color denoting the energy of the electron. b) spatial distribution of the energy deposited distribution in water.

From Monte Carlo simulations, it is clear that the dose varies dramatically as a function of radial distance from the point of incidence of the primary beam. Since most experiments are done in scanning mode, literature typically reports the dose at the surface, averaged over the pixel area, in units of e⁻/nm², referred to as $c_p^o$, or in units of Gy (J/kg), referred to as $\varphi_p^o$ in the main text. The dose can be expressed as

$$c_p^o = \frac{I_B \, \tau_D \, n}{A_p} \qquad (3)$$

Where $I_B$ is the beam current in e⁻/ sec, $A_p$ is the pixel area in nm², $\tau_D$ is the dwell-time per scan in seconds and n is the number of scans. Unless otherwise mentioned, in this study the pixel size is 100 nm x 100 nm and the number of scans is 1.

This can be further converted into pixel averaged surface dose in Gy (J/kg) using:

$$\varphi_p^o \left(Gy \text{ or } \frac{J}{kg}\right) = S_P \left(MeV \frac{cm^2}{g}\right) \left(\frac{1.6 * 10^{-19} * 10^6}{10^{-3}}\right) * c_p^o \left(\frac{\# \text{ of electrons}}{nm^2}\right) \left(10^{14} \frac{nm^2}{cm^2}\right) \qquad (4)$$

here $S_P$ is the density normalized stopping power for electrons, which in turn is a function of the energy of electrons. For example, once we know the $c_p^o$ for the set of parameters (1nA, 1 ms 100nm x 100nm), for known energy of primary beam (say $E = 3$ keV, with the $S_P$ of electrons in water in 56.21 MeV cm²/g), $\varphi_p^o$ is 5.61 x 10⁸ Gy.

Table S1 A few examples of the dose values used for PEG crosslinking and electron microscopy of microorganisms.

| Radiation & parameters | Crosslinking dose & conversion | Comment | Reference |
|---|---|---|---|
| X-rays ~12 keV | ~ 3 10$^7$ Gy | Pegilation of Au NP in solution | 50 |
| Electrons 10 keV 20-100 pA | ~ 0.1 C/m2= 1 e/nm2 | Dry PEG 6800 | 23 |
| 50 eV electrons | ~ 5·10$^3$ e/nm$^2$ | Carbonization of PEG | 26 |
| | **Cells viability dose** | | |
| TEM | 0.005 $e^-$/nm$^2$ at 100 kV | Reproductive death of E. coli | 51 |
| TEM | 37 $e^-$/nm$^2$, | Minimum dose required for high resolution (5 nm) imaging of bio specimen (note that this is larger than viability dose) | 51 pp 468–480 |
| TEM | 6.2 × 10$^{-4}$ $e^-$/nm$^2$ | colony-forming properties of *E. coli* | Isaacson, M. S. Specimen Damage in the Electron Microscope. In 52 pp 43–44 |
| TEM | 1 to 80 $e^-$/nm$^2$ | E.coli increasingly compromised after ca 30 e/nm2 | 42 |
| ESEM 30 kV | 10$^3$ to 10$^5$ e$^-$/ nm$^2$ | fixed COS7 fibroblasts, can be kept undamaged | 53 |

For clarity and direct comparison with literature values, we report the $c_p^o$ and $\varphi_p^o$ value for various instances in the main text. For more accurate estimations where the spatial distribution is needed (for example, as inputs into the Kinetic model below), we use Monte Carlo simulations.

A systematic discrepancy is observed between the experimental height and the one predicted from Monte-Carlo simulations, as shown in Figure S4

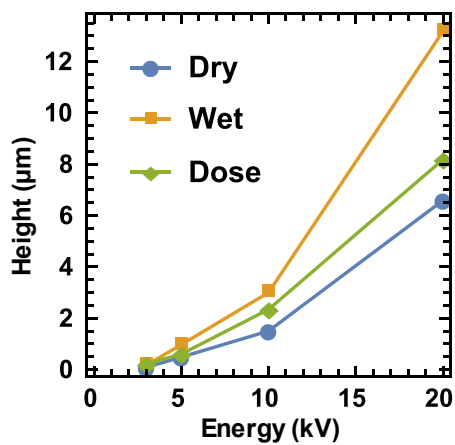

Figure S4 Height vs. Energy of Primary beam for parameters of 400 pA current and 1 ms dwell time for the electron beam. (blue) dry height measured from experiments. (orange) wet height estimated assuming 50% shrinkage. (green) Height estimated assuming critical crosslinking dose of $10^6$ Gy.

We hypothesize that this discrepancy is a result of diffusion of radiolytic species which contributes to an increase in the size of the experimental features. A kinetic model, taking into account the effect of diffusion, is therefore formulated and presented here to bolster this theory.

**Kinetic Model**

A kinetic model involving generation, reaction, and diffusion of the radiolytic species is built for application to liquids in SEM. This is based on a prior model by Schneider et al. that was developed for TEM[36,37]. The model is adapted to account for the highly non-uniform spatial dose deposition in case of SEM, by coupling it with Monte-Carlo simulations. The model framework comprises of a coupled differential equation (Eqn. 5) based on transport of dilute species for each primary and secondary radiolytic species. All parameters including the rate constants and diffusion coefficients can be found elsewhere [36].

Briefly, the model can be described as follows. Energy is deposited by the electron beam into the hydrogel solution. The calculated 2D axisymmetric energy distribution, shown in Figure S3 b, is fed as input into the kinetic model. This energy dose acts as a source for generation of primary radiolytic species via breakdown of water ($e_h^-, H, H_2, OH, H_2O_2, HO_2, H^+, OH^-$) which in turn react to produce secondary species ($HO_2^-, HO_3, O_2, O_2^-, O_3, O_3^-, O^-$). Empirical G-values[36,37] for primary radiolytic species (

Table S) are used to correlate dose and the concentration of radiolytic species produced (Eq. 6). G-values for secondary species is 0. All species are allowed to react and diffuse until they reach a steady state (~1 ms).

Table S2 G-values for primary radiolytic species

| Species | G-values (molecules/100 eV) |
|---|---|
| $e_h^-$ | 3.47 |
| H | 1.00 |
| $H_2$ | 0.17 |
| OH | 3.63 |
| $H_2O_2$ | 0.47 |
| $HO_2$ | 0.08 |
| $H^+$ | 4.42 |
| OH⁻ | 0.95 |

$$\frac{dC_i(r,z)}{dt} = D_i \nabla^2 C_i(r,z) + \sum_{j,k \neq i} k_{jk} C_j(r,z) C_j(r,z) - \sum_l k_{il} C_i(r,z) C_l(r,z) + S_i(r,z) \quad (5)$$

$$S_i(r,z) = \varphi_B(r,z) G_i \quad (6)$$

Where $(r,z)$ are cylindrical coordinates, axisymmetric across the vertical axis along the line of incidence of the Primary beam. $C_i$ is the concentration, $D_i$ is the diffusion constant, $S_i$ is the source term, $\varphi_B$ is the energy density deposited and $G_i$ is the G-value of the $i^{th}$ species.

**Imaging of composite gels with SEM**

For many applications, it is important to image the encapsulated particles inside the gel with high spatial resolution. Figure S5 depicts the SEM image of 50 nm Au nanoparticles entrapped inside the crosslinked hydrogel matrix collected with the detector sensitive to fast backscattered electrons (BSE). In this SEM imaging mode, the contrast of the objects is determined by the difference of the atomic numbers (Z) of the nanoparticle and matrix material as well as on the depth at which the electrons are collected. In the SEM image Au particles with much larger effective Z compared to hydrogel matrix appear brighter, and both: their signal strength and the resolution wanes with the depth of the nanoparticle inside the gel (Figure S5). To evaluate the feasible imaging depth for hydrogel embedded objects we conducted MC simulation of the BSE images of heterogeneous and compared them with the experimental data (Figure S5). As can be seen, SEM can be used to probe nanoparticles as deep as a few hundred nanometers using a 20kV beam energy with resolution still better compared to conventional optical microscopy.

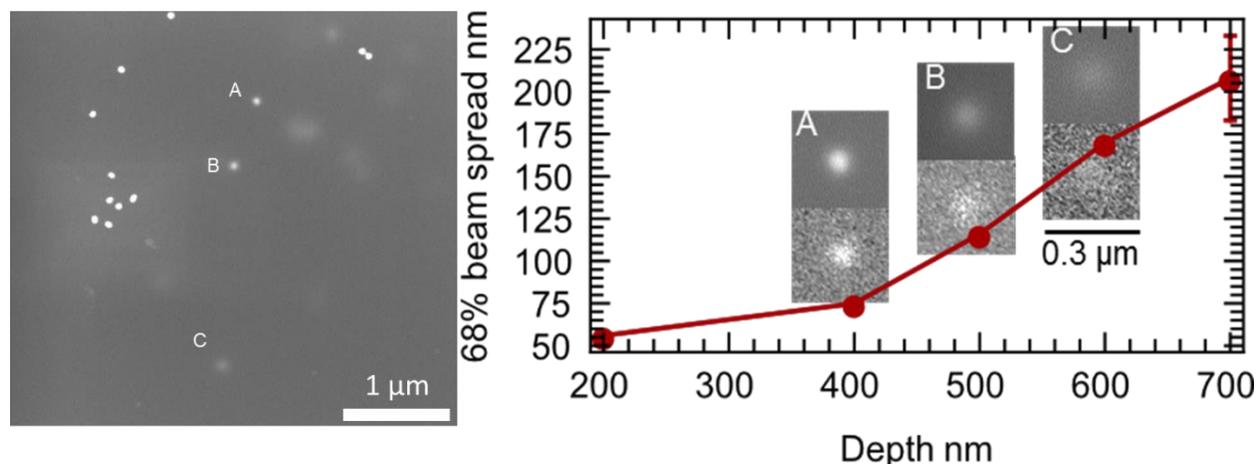

Figure S5 (Left) SEM image of embedded Au nanoparticles collected with backscattered electrons. The observed effective diameters increase, and BSE signal reduction is due to the different depth of embedded nanoparticles. (Right) MC simulated diameter of 50 nm nanoparticles as a function of the particle depth. Insets are comparisons experimentally observed and MC simulated SEM images

Raman Analysis

Raman analysis was done to validate the chemical effect of the curing process. Raman spectra of dried PEGDA correlates well with the previous studies[54,55] showing C-H-C bending peak at 1470 cm⁻¹ ( Figure S peak b) and C=C peaks at 1640 and 1410 cm⁻¹ ( Figure S peaks a, d and e). Peaks at 1600 and 1676 cm⁻¹ are from the initiator Irgacure 2959 as indicated by their appearance only after the initiator is added. Literature suggests 1600 cm⁻¹ peak corresponds to C=C stretching in an aromatic ring. Post-exposure to electron beam the C=C stretching peaks of PEGDA and unidentified peak at 1705 cm$^{-1}$ are reduced (peaks a, d and e in S6), supporting crosslinking induced by a breakdown of pi bonds.

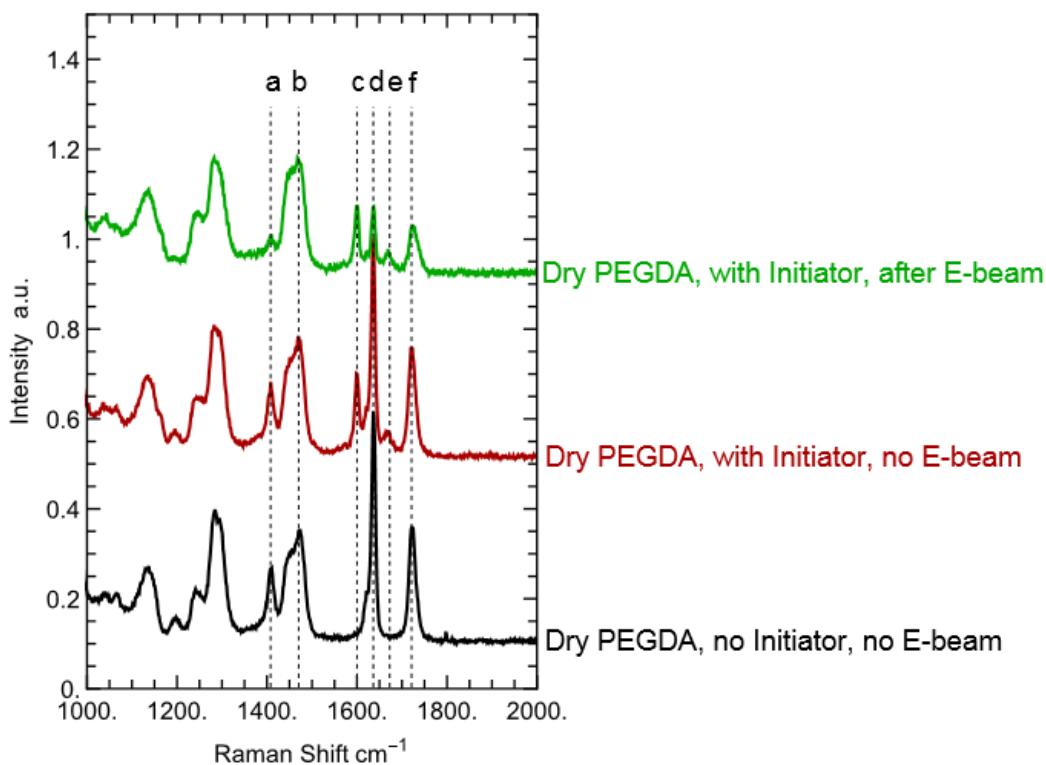

Figure S6 Raman spectra for dried PEGDA sample excited using 532 nm laser, *(black)* without initiator before e-beam exposure, *(red)* with initiator before e-beam exposure, *(green)* with initiator after e-beam exposure. All spectra are normalized w.r.t. peak b representing C-H-C bending. Peaks a, c, d, e, and f are the ones which are useful in interpreting the chemical structure and its changes on exposure to e-beam and are discussed in the text.